\documentclass[12pt,a4paper]{article}
\usepackage{bm}
\usepackage{color}
\usepackage{amsmath}
\usepackage{amsthm}
\usepackage{amssymb}
\usepackage{graphicx}
\usepackage{verbatim}
\usepackage[breaklinks]{hyperref}
\usepackage{multirow}

\newtheorem{lemma}{Lemma}
\newtheorem{theorem}{Theorem}
\newtheorem{corollary}{Corollary}

\newtheorem*{claim*}{Claim}
\newtheorem{proposition}{Proposition}

\theoremstyle{definition}
\newtheorem{definition}{Definition}

\newcommand{\set}[2]{\ensuremath{ \{ \, #1 \mid #2 \, \} }}

\renewcommand{\emptyset}{\varnothing}
\renewcommand{\epsilon}{\varepsilon}

\begin{document}
\sloppy

\title{The maximum length of shortest accepted strings for direction-determinate two-way finite automata}
\author{Olga Martynova \and Alexander Okhotin}
\maketitle

\begin{abstract}
It is shown that, for every $n \geqslant 2$,
the maximum length of the shortest string accepted
by an $n$-state direction-determinate two-way finite automaton
is exactly $\binom{n}{\lfloor\frac{n}{2}\rfloor}-1$
(direction-determinate automata are those that always remember in the current state
whether the last move was to the left or to the right).
For two-way finite automata of the general form,
a family of $n$-state automata
with shortest accepted strings of length $\frac{3}{4} \cdot 2^n - 1$ is constructed.
\end{abstract}

\section{Introduction}

A natural question about automata and related models of computation
is the length of the shortest string an automaton accepts.
A function mapping the size of an automaton
to the maximum length of the shortest accepted string,
with the maximum taken over all automata of that size,
is a certain complexity measure for a family of automata.

For one-way finite automata, this measure is trivial:
the length of the shortest string
accepted by a nondeterministic finite automaton (NFA) with $n$ states
is at most $n-1$: this is the length of the shortest path to an accepting state.
On the other hand, Ellul et al.~\cite{EllulKrawetzShallitWang}
proved that the length of shortest strings \emph{not} accepted by an $n$-state NFA
is exponential in $n$.
Similar questions were studied for other models and some variants of the problem.
Chistikov et al.~\cite{ChistikovCzerwinskiHofmanPilipczukWehar}
investigated the length of shortest strings in counter automata.
The length of shortest strings in formal grammars
under intersections with regular languages
was studied by Pierre~\cite{Pierre},
and recently by Shemetova et al.~\cite{ShemetovaOkhotinGrigorev}.
Alpoge et al.~\cite{AlpogeAngSchaefferShallit} investigated shortest strings
in intersections of deterministic one-way finite automata (DFA).

The maximum length of shortest strings
for deterministic two-way finite automata (2DFA)
has been investigated in two recent papers.
First of all, from the well-known proof
of the PSPACE-completeness of the emptiness problem for 2DFA
by Kozen~\cite{Kozen}
it is understood that the length of the shortest string
accepted by an $n$-state 2DFA
can be exponential in $n$.
There is also an exponential upper bound on this length,
given by transforming a 2DFA to an NFA:
the construction by Kapoutsis~\cite{Kapoutsis}
uses at most $\binom{2n}{n+1}=\Theta(\frac{1}{\sqrt{n}} 4^n)$ states,
and hence the length of the shortest string is slightly less than $4^n$.
Overall, the maximum length of the shortest string is exponential,
with the base bounded by 4.

The first attempt to determine the exact base
was made by Dobronravov et al.~\cite{two_way_dfa_shortest},
who constructed a family of $n$-state 2DFA
with shortest strings of length $\Omega((\sqrt[5]{10})^n) \geqslant \Omega(1.584^n)$.
The automata they have actually constructed
belong to a special class of 2DFA:
the \emph{direction-determinate automata}.
These are 2DFA with the set of states
split into states accessible only by transitions from the right
and states accessible only by transitions from the left:
in other words, direction-determinate automata always remember the direction
of the last transition in their state.

Later, Krymski and Okhotin~\cite{KrymskiOkhotin_conf}
extended the method of Dobronravov et al.~\cite{two_way_dfa_shortest}
to produce automata of a more general form, with longer shortest accepted strings.
They constructed a family of non-direction-determinate 2DFA
with shortest strings of length $\Omega((\sqrt[4]{7})^n) \geqslant \Omega(1.626^n)$.

This paper improves these bounds.
First, the maximum length of the shortest string
accepted by $n$-state direction-determinate 2DFA
is determined precisely as $\binom{n}{\lfloor\frac{n}{2}\rfloor}-1 = \Theta(\frac{1}{\sqrt{n}} 2^n)$.
The upper bound on the length of the shortest string
immediately follows from the complexity of transforming direction-determinate 2DFA to NFA,
see Geffert and Okhotin~\cite{GeffertOkhotin}.
A matching lower bound is proved by a direct construction of a family of $n$-state automata.

The second result of this paper
is that not remembering the direction helps to accept longer shortest strings:
a family of $n$-state non-direction-determinate automata
with shortest strings of length $\frac{3}{4} \cdot 2^n - 1$
is constructed.
This is more than what is possible in direction-determinate automata.

\section{Definitions}

\begin{definition}
A \emph{two-way deterministic finite automaton} (2DFA)
is a quintuple
$\mathcal{A}=(\Sigma, Q, q_0, \delta, F)$,
in which:
\begin{itemize}
\item
	$\Sigma$ is a finite alphabet,
	which does not contain two special symbols:
	the left end-marker ($\vdash$)
	and the right end-marker ($\dashv$);
\item
	$Q$ is a finite set of states;
\item
	$q_0 \in Q$ is the initial state;
\item
	$\delta \colon Q \times (\Sigma \cup \{{\vdash},{\dashv}\}) \to Q \times \{-1,+1\}$
	is a partial transition function;
\item
	$F \subseteq Q$
	is the set of accepting states,
	effective at the right end-marker ($\dashv$).
\end{itemize}
An input string $w = a_1 \ldots a_m \in \Sigma^*$ is given to an automaton
on a tape ${\vdash} a_1 \ldots a_m {\dashv}$.
The automaton starts at the left end-marker ${\vdash}$ in the state $q_0$.
At each moment, if the automaton is in a state $q \in Q$
and sees a symbol $a \in \Sigma \cup \{{\vdash}, {\dashv}\}$,
then, according to the transition function $\delta(q, a)=(r, d)$,
it enters a new state $r$
and moves to the left or to the right depending on the direction $d$.
If the requested value $\delta(q, a)$ is not defined, then the automaton rejects.
The automaton accepts the string, if it ever comes to the right end-marker $\dashv$
in any state from $F$.
The automaton can also loop.

The language recognized by an automaton $A$, 
denoted by $L(A)$, is the set of all strings it accepts.
\end{definition}

This paper also uses a subclass of 2DFA,
in which one can determine the direction of the previous transition
from the current state.

\begin{definition}[\cite{KuncOkhotin_reversible}]
A 2DFA is called \emph{direction-determinate}, 
if there is a partition of the set of states $Q=Q^+ \cup Q^-$,
with $Q^+ \cap Q^- = \emptyset$,
such that for each transition $\delta(q, a)=(r, +1)$, the state $r$ must belong to $Q^+$,
and for each transition $\delta(q, a)=(r, -1)$, the state  $r$ is in $Q^-$.
\end{definition}

The known upper bounds on the length of the shortest accepted string
are different for direction-determinate 2DFA and for 2DFA of the general form.
These bounds are inferred from the complexity of transforming
two-way automata with $n$ states to one-way NFA:
for 2DFA of the general form,
as proved by Kapoutsis~\cite{Kapoutsis},
it is sufficient and in the worst case necessary
to use $\binom{2n}{n}$ states in a simulating NFA,
whereas for direction-determinate 2DFA
the simulating 2DFA requires $\binom{n}{\lfloor\frac{n}{2}\rfloor}$ states in the worst case,
see Geffert and Okhotin~\cite{GeffertOkhotin}.
Since the shortest string in a language cannot be longer
than the shortest path to an accepting state in an NFA,
the following bounds hold.

\begin{theorem}[Dobronravov et al.~\cite{two_way_dfa_shortest}]
Let $n \geqslant 1$,
and let $A$ be a 2DFA with $n$ states,
which accepts at least one string.
Then the length of the shortest string accepted by $A$
is at most $\binom{2n}{n}-1$.
If the automaton $A$ is direction-determinate,
then the length of the shortest accepted string
does not exceed $\binom{n}{\lfloor\frac{n}{2}\rfloor}-1$.
\end{theorem}

The first result of this paper is that
this upper bound for direction-determinate automata
is actually precise.

\section{Shortest accepted strings for direction-determinate automata}

In this section, direction-determinate automata
with the maximum possible length $\binom{n}{\lfloor\frac{n}{2}\rfloor}-1$
of shortest accepted strings,
where $n$ is the number of states,
will be constructed.

Automata are constructed for every $k$ and $\ell$,
where $k$ is the number of states reachable by transitions to the right
and $\ell$ is the number of states reachable in the left direction.
The following theorem shall be proved.

\begin{theorem}\label{dirdet_shortest_theorem}
For every $k \geqslant 2$ and $\ell \geqslant 0$
there exists a direction-determinate 2DFA with the set of states $Q=Q^+ \cup Q^-$,
where $|Q^+|=k$ and $|Q^-|=\ell$,
such that the length of the shortest string it accepts is $\binom{k+\ell}{\ell+1}-1$.
\end{theorem}

The automaton constructed in the theorem works as follows.
While working on its shortest string,
it processes every pair of consecutive symbols
by moving back and forth between them,
thus effectively comparing them to each other.
Eventually it moves on to the next pair and processes it in the same way.
It cannot come back to the previous pair anymore,
because it has no transitions for that.

The automaton's motion between two neighbouring symbols
begins when it first arrives from the first symbol to the second in some state from $Q^+$.
Then it moves back and forth,
alternating between states from $Q^+$ at the second symbol
and states from $Q^-$ at the first symbol,
and finally leaves the second symbol to the right.
Among the states visited by the automaton during this back-and-forth motion,
the number of states from $Q^+$ is greater by one than the number of states from $Q^-$.
Two such sets of states will be denoted
by a pair $(P, R)$, where $P \subseteq Q^-$, $R \subseteq Q^+$ and $|R|=|P|+1$.

\begin{proposition}
There are $\binom{k+\ell}{\ell+1}$
different pairs $(P, R)$, such that $P \subseteq Q^-$, $R \subseteq Q^+$ and $|R|=|P|+1$.
\end{proposition}
\begin{proof}
There are as many pairs $(P, R)$ as pairs $(Q^- \setminus P, R)$, where $|R|=|P|+1$.
The number of pairs of the latter form
is equal to the number of subsets of $Q$ of size $\ell+1$,
that is, $\binom{k+\ell}{\ell+1}$.
\end{proof}

Let the sets $Q^+$ and $Q^-$ be linearly ordered.
Then one can define an order on the set of pairs $(P, R)$ as follows.
In every such pair,
let $P=\{p_1, \ldots, p_m\}$, where $p_1 < \ldots < p_m$,
and $R=\{r_1, \ldots, r_{m+1}\}$, where $r_1 < \ldots < r_{m+1}$.
There is a corresponding sequence to each pair,
of the form
$r_1$, $-p_1$, $r_2$, $-p_2$, \ldots, $r_m$, $-p_m$, $r_{m+1}$,
and different pairs are compared by the lexicographic order on these sequences.
In Table~\ref{t:order_on_pairs_P_R},
all pairs $(P, R)$, for $k=4$ and $\ell=2$,
are given in increasing order,
along with the corresponding sequences.

\begin{table}[t]
\begin{equation*}
\begin{array}{ll}
\text{pairs } (P, R) & \text{sequences} \\
\hline
\emptyset, \{1\} &  (1)\\
\{2'\}, \{1, 2\} &  (1,-2',2)\\
\{2'\}, \{1, 3\} &  (1,-2',3)\\
\{2'\}, \{1, 4\} &  (1,-2',4)\\
\{1'\}, \{1, 2\} &  (1,-1',2)\\
\{1', 2'\}, \{1, 2, 3\} &  (1,-1',2,-2',3)\\
\{1', 2'\}, \{1, 2, 4\} &  (1,-1',2,-2',4)\\
\{1'\}, \{1, 3\} &  (1,-1',3)\\
\{1', 2'\}, \{1, 3, 4\} &  (1,-1',3,-2',4)\\
\{1'\}, \{1, 4\} &  (1,-1',4)\\
\emptyset, \{2\} &  (2)\\
\{2'\}, \{2, 3\} &  (2,-2',3)\\
\{2'\}, \{2, 4\} &  (2,-2',4)\\
\{1'\}, \{2, 3\} &  (2,-1',3)\\
\{1', 2'\}, \{2, 3, 4\} &  (2,-1',3,-2',4)\\
\{1'\}, \{2, 4\} &  (2,-1',4)\\
\emptyset, \{3\} &  (3)\\
\{2'\}, \{3, 4\} &  (3,-2', 4)\\
\{1'\}, \{3, 4\} &  (3,-1',4)\\
\emptyset, \{4\} &  (4)
\end{array}
\end{equation*}
\caption{All pairs $(P, R)$
for sets of states $Q^+=\{1, 2, 3, 4\}$ and $Q^-=\{1', 2'\}$.}
\label{t:order_on_pairs_P_R}
\end{table}

Let $N=\binom{k+\ell}{\ell+1}$ be the number of pairs.
Then all pairs are enumerated in increasing order
as 
$(P^{(1)}, R^{(1)}) < \ldots < (P^{(N)}, R^{(N)})$,
where
$P^{(i)}=\{p^{(i)}_1, \ldots, p^{(i)}_{m_i}\}$ and
$R^{(i)}=\{r^{(i)}_1, \ldots, r^{(i)}_{m_i+1}\}$.
In particular, the least pair is $(P^{(1)}, R^{(1)})=(\emptyset, \{\min Q^+\})$,
because the corresponding sequence ($\min Q^+$) is lexicographically the least.
The greatest pair is $(P^{(N)}, R^{(N)})=(\emptyset, \{\max Q^+\})$.

The desired direction-determinate automaton $A$
with the shortest accepted string of length $N-1$
is defined over an alphabet $\Sigma=\{a_1, \ldots, a_{N-1}\}$,
and the shortest accepted string will be $w=a_1 \ldots a_{N-1}$.
The set of states is defined as $Q=Q^+ \cup Q^-$,
where $Q^+=\{1, \ldots, k\}$ and $Q^-=\{1', \ldots, \ell'\}$.
The initial state is $q_0=1$.
The only transition by the left end-marker ($\vdash$)
leads from the initial state to the least state in $R^{(1)}$.
\begin{subequations}
\begin{align}
	\label{A_transition_initial}
	\delta(q_0, {\vdash}) &= (r^{(1)}_1, +1)
\intertext{%
For each symbol $a_i$, transitions are defined in the states $R^{(i)} \cup P^{(i+1)}$.
If the automaton is at the symbol $a_i$ in any state from $R^{(i)}$ (except for the greatest state),
then it moves to the left in the corresponding state from $P^{(i)}$.
}
	\label{A_transition_r_to_p}
	\delta(r^{(i)}_j, a_i) &= (p^{(i)}_j, -1)
		&& (j \in \{1, \ldots, m_i\})
\intertext{%
For the greatest state in $R^{(i)}$,
there is no corresponding state in $P^{(i)}$,
and so the automaton moves to the right
(and this is the only way to move from $Q^+$ to $Q^+$,
and hence the only way to advance from the symbol $a_i$ to the next symbol for the first time).
}
	\label{A_transition_r_to_next_r}
	\delta(r^{(i)}_{m_i+1}, a_i) &= (r^{(i+1)}_1, +1)
\intertext{%
In each state from $P^{(i)}$,
the automaton moves to the right
in the next available state from $R^{(i)}$.
}
	\label{A_transition_p_to_r}
	\delta(p^{(i+1)}_j, a_i) &= (r^{(i+1)}_{j+1}, +1)
		&& (j \in \{1, \ldots, m_{i+1}\})
\end{align}
\end{subequations}
There are no transitions at the right end-marker,
and there is one accepting state: $F=\{r^{(N)}_{m_N+1}\}$.

\begin{figure}[t]
	\centerline{\includegraphics[scale=1]{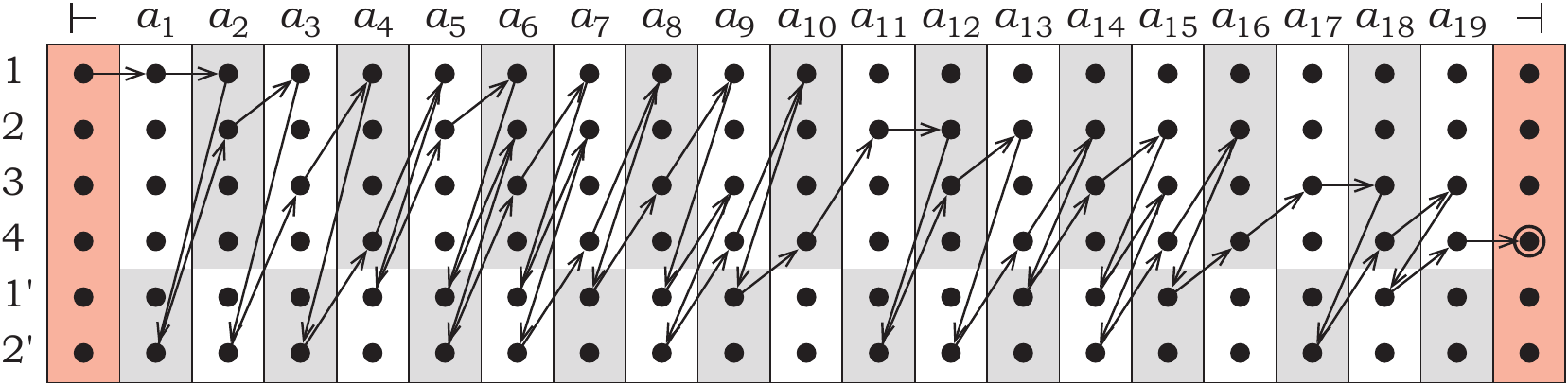}}
	\caption{The accepting computation of the automaton $A$ on the string $w$,
		for $k=4$ and $\ell=2$.}
	\label{f:example_k4_l2}
\end{figure}

The computation of the automaton on the string $w = a_1 \ldots a_{N-1}$
is illustrated in Figure~\ref{f:example_k4_l2}.
The automaton gradually advances,
and moves between every two subsequent symbols,
$a_{i-1}$ and $a_i$,
according to the sets $P_i$ and $R_i$.
Transitions at $a_i$ expect that there is $a_{i-1}$ to the left,
whereas transitions at $a_{i-1}$ expect $a_i$ to the right.
As long as every symbol is followed by the next symbol in order,
these expectations will be fulfilled each time,
and the automaton accepts in the end.

\begin{lemma}
The automaton $A$ accepts the string $w=a_1 \ldots a_{N-1}$.
\end{lemma}
\begin{proof}
It is claimed that the automaton $A$, executed on the string $w$,
eventually arrives to each symbol $a_i$ in the state $r^{(i)}_{m_i + 1}$.
This is proved by induction on $i$.

Base case $i=1$:
the first transition \eqref{A_transition_initial}
moves the automaton to the state $r^{(1)}_1$.
The first pair $(P^{(1)}, R^{(1)})$ is $(\emptyset, \{1\})$,
and so $r^{(1)}_1 = r^{(1)}_{m_1+1}$.

Induction step.
Assume that the automaton comes to the symbol $a_i$ in the state $r^{(i)}_{m_i+1}$.
Then it makes a transition~\eqref{A_transition_r_to_next_r} to the right in the state $r^{(i+1)}_1$.
Then it executes the sequence of transitions \eqref{A_transition_r_to_p}, \eqref{A_transition_p_to_r},
defined by the pair $(P_{i+1}, R_{i+1})$,
moving back and forth between $a_{i+1}$ and $a_i$,
and passing through the states $p^{(i+1)}_1$, $r^{(i+1)}_2$, $p^{(i+1)}_2$, \ldots 
$r^{(i+1)}_{m_{i+1}}$, $p^{(i+1)}_{m_{i+1}}$, $r^{(i+1)}_{m_{i+1}+1}$.
And so it comes to the symbol $a_{i+1}$ in the state $r^{(i+1)}_{m_{i+1}+1}$,
as shown in Figure~\ref{f:dirdet_shortest_proof}.

\begin{figure}[t]
	\centerline{\includegraphics[scale=1]{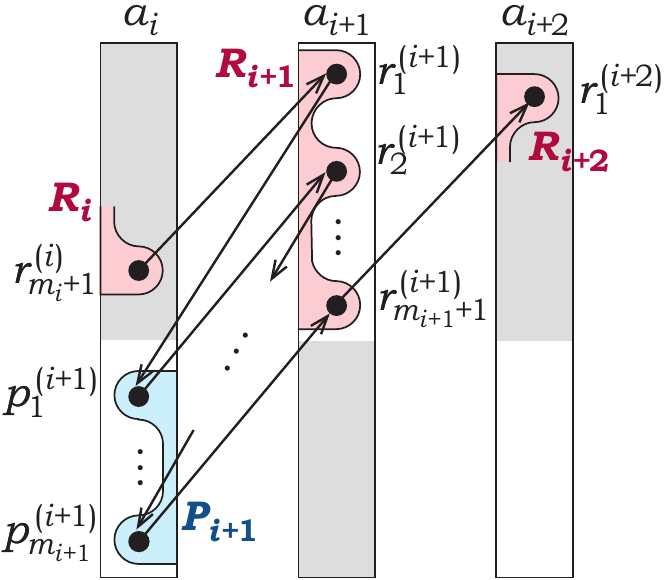}}
	\caption{The moves of $A$ between two neighbouring symbols of $w$.}
	\label{f:dirdet_shortest_proof}
\end{figure}

In the end, the automaton comes to the last symbol $a_{N-1}$ in the state $r^{(N-1)}_{m_{N-1}+1}$.
Then it makes a transition~\eqref{A_transition_r_to_next_r}
and moves to the right end-marker in the state $r^{(N)}_1$.
And this is the accepting state $r^{(N)}_{m_N+1}$,
because the last pair $(P^{(N)}, R^{(N)})$ is $(\emptyset, \{k\})$.
Therefore, the string $w$ is accepted.
\end{proof}

It is claimed that the automaton $A$ cannot accept any shorter string.
It cannot accept the empty string;
if it did, then the first transition would lead to the right end-marker in the state 1,
and the automaton would reject, because $k \neq 1$.
Next, it will be shown that each accepted string
begins with the symbol $a_1$
and ends with the symbol $a_{N-1}$.
Finally, it will be proved that the automaton cannot skip any number,
that is, the number of every next symbol,
as compared to the number of the previous symbol,
cannot increase by more than $1$.
If the number decreases or does not change, this would make the string only longer;
but in order to reach $a_{N-1}$ from $a_1$ without skipping any number,
the automaton would have to move through all symbols of the alphabet,
and therefore an accepted string cannot be shorter than $N-1$ symbols.

\begin{lemma}
Every string accepted by the automaton $A$
begins with the symbol $a_1$.
\end{lemma}
\begin{proof}
Let the automaton $A$ accept some string
that starts from some symbol $a_i$.
The transition from the initial configuration
leads the automaton to the state $r^{(1)}_1$ at the first symbol $a_i$.
As $(P^{(1)}, R^{(1)}) = (\emptyset, \{1\})$,
the state $r^{(1)}_1$ is $1$.

Transitions by the symbol $a_i$ are defined only in states from $R^{(i)} \cup P^{(i+1)}$,
and hence $1 \in R^{(i)}$, for otherwise the automaton immediately rejects.
If there is at least one more state in $R^{(i)}$,
then the transition in the state $1$ by $a_i$ moves the automaton to the left.
Then the automaton returns to the left end-marker,
and then either loops or rejects,
because there is only one transition defined there.
Therefore, there are no other states in $R^{(i)}$ besides $1$,
and so, $(P^{(i)}, R^{(i)})=(\emptyset, \{1\})=(P^{(1)}, R^{(1)})$,
which implies $i=1$.
\end{proof}

\begin{lemma}
Every string accepted by the automaton $A$
ends with the symbol $a_{N-1}$.
\end{lemma}
\begin{proof}
Let a string accepted by $A$ end with a symbol $a_i$.
To accept, the automaton should move from $a_i$ to the right
using the transition~\eqref{A_transition_r_to_next_r},
and it arrives to the right end-marker in the state $r^{(i+1)}_1$.
As the only accepting state is $k$,
and the automaton rejects at the right end-marker in all other states,
this state must be $r^{(i+1)}_1=k$.
Because the state $r^{(i+1)}_1$ is the least in $R^{(i+1)}$,
it follows that $R^{(i+1)}=\{k\}$ and $P^{(i+1)}=\emptyset$.
Therefore, this is the last pair, and $i=N-1$.
\end{proof}

\begin{lemma}
No string accepted by the automaton $A$
may contain any substring of the form $a_i a_j$, where $j>i+1$.
\end{lemma}
\begin{proof}
The proof is by contradiction.
Suppose that $A$ accepts a string
that contains a substring $a_i a_j$, with $j>i+1$.
In order to accept,
the automaton should eventually reach this symbol $a_j$ for the first time,
moving to it from the symbol $a_i$.
To make this transition, the automaton should be at $a_i$ in some state from $Q^+$
(indeed, if it were in the state from $Q^-$,
then it would have been at $a_j$ already at the previous step).
Then the automaton must use the transition~\eqref{A_transition_r_to_next_r}
to move from $a_i$ to $a_j$,
and this transition leads to the state $r^{(i+1)}_1$.
For the computation to go onward,
this state should lie in $R^{(j)}$.
Moreover, the state $r^{(i+1)}_1$ should be the least in $R^{(j)}$,
for otherwise the pair $(P^{(j)}, R^{(j)})$
would be less than the pair $(P^{(i+1)}, R^{(i+1)})$.
Also $r^{(i+1)}_1$ cannot be the only state in $R^{(j)}$:
if not, then $(P^{(j)}, R^{(j)})$ would either coincide with or be less than $(P^{(i+1)}, R^{(i+1)})$.

It can be concluded that $r^{(i+1)}_1=r^{(j)}_1$,
and the next transition from this state leads to the state $p^{(j)}_1$,
moving to the symbol $a_i$.
For the automaton to have a transition in the state $p^{(j)}_1$ at $a_i$,
this state should belong to $P^{(i+1)}$.
In addition, it should be the least among the state in $P^{(i+1)}$,
because if there were a lesser state $p$,
then the second term in the sequence for $(P^{(i+1)}, R^{(i+1)})$ would be $-p$,
and this pair would be greater than $(P^{(j)}, R^{(j)})$.
This leads to the equality $p^{(j)}_1=p^{(i+1)}_1$.

By analogous arguments, one can prove that
the sequences for $(P^{(j)}, R^{(j)})$ and for $(P^{(i+1)}, R^{(i+1)})$
must coincide and continue infinitely.
This is impossible, because the numbers of states increase,
and there finitely many of them.
\end{proof}

\begin{corollary}[from Theorem~\ref{dirdet_shortest_theorem}]
For every $n \geqslant 2$,
there is a direction-determinate 2DFA with $n$ states,
such that the length of the shortest string it accepts is $\binom{n}{\lfloor\frac{n}{2}\rfloor}-1$.
\end{corollary}

\section{Longer shortest strings for automata of the general form}

\begin{figure}[t]
	\centerline{\includegraphics[scale=1]{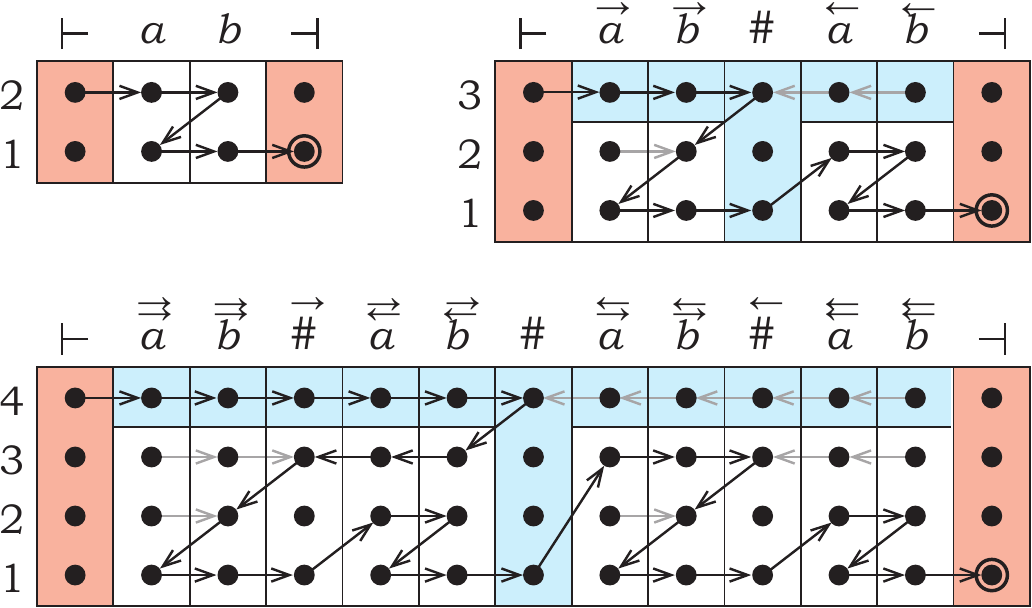}}
	\caption{Computations of automata $A_2$, $A_3$ and $A_4$
		from the proof of Theorem~\ref{theorem_2_pow_n}
		on their shortest strings $w_2$, $w_3$ and $w_4$.}
	\label{f:2dfa_shortest_2_3_4}
\end{figure}

The main result of this section is the construction of a family of 2DFA
with shortest strings of length $3\cdot 2^{n-2}-1$,
where $n$ is the number of states in an automaton.
This is more than the maximum possible length of shortest strings
for direction-determinate automata;
in other words, \emph{forgetting the direction is useful.}

\begin{theorem}\label{theorem_2_pow_n}
For each $n \geqslant 2$ there exists a 2DFA with $n$ states,
such that the shortest string it accepts is of length $3\cdot 2^{n-2}-1$.
\end{theorem}
\begin{proof}
The automata and the shortest strings they accept
are constructed inductively;
for small values of $n$ they are given in Figure~\ref{f:2dfa_shortest_2_3_4}.

For the inductive proof to work,
the following set of properties is ensured for every $n$.

\begin{claim*}
For each $n \geqslant 2$ there exists a 2DFA $A_n = (\Sigma_n, Q_n, \delta_n)$
with no transitions by end-markers, no initial state and no accepting states,
with the set of states $Q_n = \{1, \ldots, n\}$,
and there exists a string $w_n \in \Sigma_n^*$
of length $3\cdot 2^{n-2}-1$,
such that the following two properties hold.
\begin{enumerate}
\item
	If $A_n$ starts at any symbol of $w_n$ in the state $n$,
	then it eventually leaves this string
	by a transition from its rightmost symbol to the right in the state $1$.
\item
	If for some non-empty string $u$ there exists a position,
	in which the automaton $A_n$ can start in the state $n$
	and eventually leave the string $u$
	by a transition from its rightmost symbol to the right in the state $1$,
	then $u$ is at least as long as $w_n$.
\end{enumerate}
\end{claim*}

The first observation is that Theorem~\ref{theorem_2_pow_n}
follows from this claim.
Let $n \geqslant 2$,
and let $A_n$ and $w_n$ be an automaton and a string
that satisfy the conditions in the claim.
Then $A_n$ is supplemented
with an initial state $n$,
a set of accepting states $\{1\}$
and a single transition by the left end-marker: from the state $n$ to the state $n$;
no transitions by the right end-marker are defined.
The resulting automaton $A'_n$ becomes a valid 2DFA,
and it accepts the string $w_n$ as follows:
from the initial state at $\vdash$
it moves to the first symbol of $w_n$ in the state $n$,
then, by the first point of the claim,
the automaton eventually leaves $w_n$ to the right in the state $1$,
and thus arrives to the right end-marker $\dashv$ in an accepting state.

To see that every string accepted by $A'_n$ is of length at least $|w_n|$,
let $u$ be any accepted string.
It is not empty, because on the empty string
the automaton steps on the right end-marker in the state $n$ and rejects.
Then, after the first step the automaton $A'_n$
is at the first symbol of $u$ in the state $n$.
It cannot return to $\vdash$,
because it has already used the only transition at this label,
and if it ever comes back, it will reject or loop.
Also the automaton cannot come to $\dashv$ in states other than $1$.
In order to accept, it must arrive to $\dashv$ in the state $1$,
and this is the first and the only time when it leaves the string $u$.
Then, by the second point of the claim,
the length of $u$ cannot be less than the length of $w_n$.

It remains to prove the claim,
which is done by induction on $n$.

Base case: $n = 2$.
 
The automaton $A_2 = (\Sigma_2, Q_2, \delta_2)$ for $n = 2$ is constructed as follows.
The alphabet is $\Sigma_2 = \{a, b\}$, and the set of states is $Q_2 = \{1, 2\}$.
The transition function is defined by
\begin{align*}
\delta_2(2,a) &= (2,+1), \\
\delta_2(2,b) &= (1,-1), \\
\delta_2(1,a) &= (1,+1), \\
\delta_2(1,b) &= (1,+1).
\end{align*}

The string $w_2$ is $ab$,
and the computation of $A_2$ on $w_2$ is presented in Figure~\ref{f:2dfa_shortest_2_3_4} (top left).
To be precise, computations starting in the state $2$ either at $a$ or at $b$
both end by leaving the string to the right in the state $1$, as claimed.
There are only two shorter non-empty strings: $a$ and $b$.
If the automaton starts on the string $a$ in the state $2$,
then it moves to the right in the state $2$;
on $b$, it moves to the left in the state $1$.
In either case, it does not go to the right in the state $1$.
Thus, the second point of the claim is satisfied.
The length of the string is $|w_2| = 2 = 3 \cdot 2^0-1$.

Induction step: $n \to n+1$.

Let an $n$-state 2DFA $A_n = (\Sigma_n, Q_n, \delta_n)$
and a string $w_n \in \Sigma_n^*$
satisfy the claim.
The $(n+1)$-state automaton $A_{n+1}$ satisfying the claim
is constructed as follows.
Let $A_{n+1}  = (\Sigma_{n+1}, Q_{n+1}, \delta_{n+1})$.
\begin{itemize}
\item	
	Its alphabet is
	$\Sigma_{n+1} = \overrightarrow{\Sigma_n} \cup \overleftarrow{\Sigma_n} \cup \{\#\}$,
	where $\overrightarrow{\Sigma_n} = \set{\overrightarrow{a}}{a \in \Sigma_n}$
	and $\overleftarrow{\Sigma_n} = \set{\overleftarrow{a}}{a \in \Sigma_n}$
\item
	The set of states is $Q_{n+1} = Q_n \cup \{n+1\} = \{1,\ldots, n+1\}$.
\item
	The transition function is defined as follows.
	In the new state $n+1$,
	the automaton moves by all symbols with arrows
	in the directions pointed by the arrows.
	\begin{align*}
		\delta_{n+1}(n+1,\overrightarrow{a}) &= (n+1,+1),
			&& \text{for } a \in \Sigma
				\\
		\delta_{n+1}(n+1,\overleftarrow{a}) &= (n+1,-1),
			&& \text{for } a \in \Sigma
	\intertext{%
	In all old states $1, \ldots, n$,
	on symbols with arrows,
	the new automaton works in the same way
	as the automaton $A_n$ on the corresponding symbols without arrows.
	}
		\delta_{n+1}(i,\overrightarrow{a}) = \delta_{n+1}(i,\overleftarrow{a}) &=
		\delta_n(i,a),
			&& \text{for } a \in \Sigma \text{ and } i \in \{1, \ldots, n\}
				\\
	\intertext{%
	By the new separator symbol $\#$,
	only two transitions are defined.
	In the state $n+1$, the automaton moves to the left in the state $n$,
	thus starting the automaton $A_n$
	on the substring to the left.
	}
		\delta_{n+1}(n+1,\#) &= (n,-1)
	\intertext{%
	And if the automaton gets to $\#$ in the state $1$
	(which happens after concluding the simulation of $A_n$
	on the substring to the left),
	then the automaton moves to the right in the state $n$
	to start the simulation of $A_n$ also on the substring to the right of the separator $\#$.
	}
		\delta_{n+1}(1,\#) &= (n,+1)
	\end{align*}
	The rest of transitions are undefined.
\end{itemize}

\begin{figure}[t]
	\centerline{\includegraphics[scale=1]{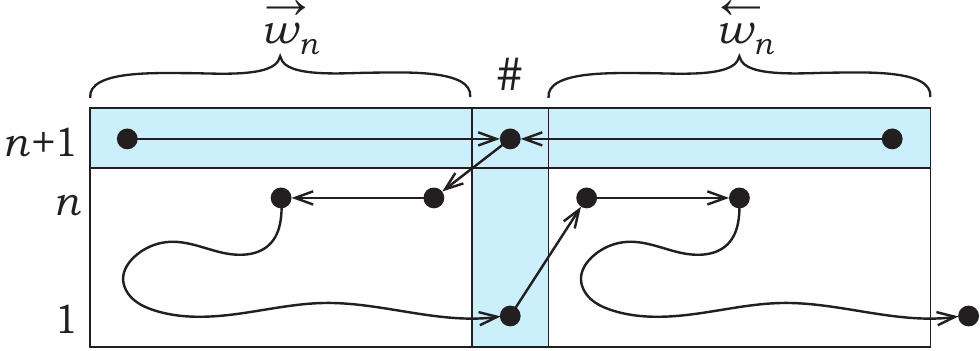}}
	\caption{Computation of the automaton $A_{n+1}$ on the string $w_{n+1}$.}
	\label{f:theorem_2_pow_n_computation_on_w_n_plus_1}
\end{figure}

Note that once the automaton $A_{n+1}$ leaves the state $n+1$,
it never returns to it,
because there are no transitions to $n+1$ from any other state.
Let
$h \colon (\overrightarrow{\Sigma_n}\cup \overleftarrow{\Sigma_n})^* \to \Sigma_n^*$
be a string homomorphism
which removes the arrow from the top of every symbol,
that is, 
$h(\overrightarrow{a})=h(\overleftarrow{a})=a$ for all $a \in \Sigma_n$.
The automaton $A_{n+1}$
works in the states $1, \ldots, n$
on symbols from $\overrightarrow{\Sigma_n}\cup \overleftarrow{\Sigma_n}$
as $A_n$ works on the corresponding symbols from $\Sigma_n$.
Then, if $h(w) = w_n$ for some $w \in (\overrightarrow{\Sigma_n}\cup \overleftarrow{\Sigma_n})^*$,
it follows that the automaton $A_{n+1}$,
having started in the state $n$ at any symbol of $w$,
eventually leaves the string $w$ by moving to the right in the state $1$.
Furthermore, if $|w| < |w_n|$
for some string $w \in (\overrightarrow{\Sigma_n}\cup \overleftarrow{\Sigma_n})^*$,
then the automaton $A_{n+1}$,
having started in the state $n$ at any symbol of $w$,
cannot leave the string by moving to the right in the state $1$.

The string $w_{n+1}$ is defined as $\overrightarrow{w_n}\#\overleftarrow{w_n}$,
where $\overrightarrow{a_1 \ldots a_\ell} = \overrightarrow{a_1} \ldots \overrightarrow{a_\ell}$
and $\overleftarrow{a_1 \ldots a_\ell} = \overleftarrow{a_1} \ldots \overleftarrow{a_\ell}$
for every string $a_1 \ldots a_\ell \in \Sigma_n^*$.
The length of $w_{n+1}$ is
$|w_{n+1}| = 2|w_n|+1 = 2(3\cdot 2^{n-2}-1)+1 = 3 \cdot 2^{n-1}-1$,
as desired.

First, it is proved that the automaton $A_{n+1}$ works on the string $w_{n+1}$
as stated in the first point of the claim.
Let $A_{n+1}$ start its computation on the string $w_{n+1}$
at any symbol in the state $n+1$,
as shown in Figure~\ref{f:theorem_2_pow_n_computation_on_w_n_plus_1}.
By the symbols in $\overrightarrow{w_n}$, the automaton moves to the right,
maintaining the state $n+1$;
by the symbols in $\overleftarrow{w_n}$, it moves to the left in $n+1$.
Thus, wherever the automaton begins,
it eventually arrives to the separator $\#$ in the state $n+1$.
Next, the automaton moves to the last symbol of $\overrightarrow{w_n}$ in the state $n$.
Since $h(\overrightarrow{w_n}) = w_n$,
the automaton $A_{n+1}$ operates on $\overrightarrow{w_n}$ as $A_n$ on $w_n$,
and leaves $\overrightarrow{w_n}$ by a transition to the right in the state $1$.
Then $A_{n+1}$ arrives to the separator $\#$ again, now in the state $1$,
and moves to the first symbol of $\overleftarrow{w_n}$ in the state $n$.
As $h(\overleftarrow{w_n}) = w_n$, the automaton $A_{n+1}$ works as $A_n$ on $w_n$,
and leaves $\overleftarrow{w_n}$ (and the whole string $w_{n+1}$)
by moving to the right in the state $1$.

Turning to the second point of the claim,
it should be proved that computations of a certain form
are impossible on any strings shorter than $w_{n+1}$.
Let $w \in \Sigma_{n+1}^*$ be a string,
and let there be a position in $w$,
such that the automaton $A_{n+1}$,
having started at this position in the state $n+1$,
eventually leaves the string $w$ by a transition to the right in the state $1$.
It is claimed that $|w| \geqslant |w_{n+1}|$.
 
Consider the computation of $A_{n+1}$
leading out of $w$ to the right in the state $1$.
It begins in the state $n+1$,
and the automaton maintains the state $n+1$ at all symbols except $\#$.
In order to reach the state $1$, there should be a moment in the computation on $w$
when the automaton arrives at some symbol $\#$ in the state $n+1$.
Let $u$ be the prefix of $w$ to the left of this $\#$,
and let $v$ be the suffix to the right of this $\#$;
note that the substrings $u$ and $v$ may contain more symbols $\#$.
It is sufficient to prove that $|u| \geqslant |w_n|$ and $|v| \geqslant |w_n|$.

\begin{figure}[t]
	\centerline{\includegraphics[scale=1]{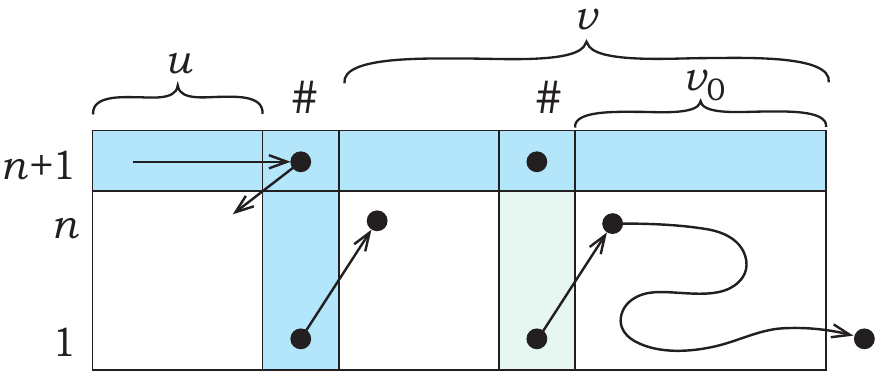}}
	\caption{The partition $w=u\#v$ and the suffix $v_0$ of $v$.}
	\label{f:theorem_2_pow_n_u3}
\end{figure}

Consider first the case of the suffix $v$.
Let $v_0$ be the longest suffix of $v$ that does not contain the symbol $\#$;
then the symbol preceding $v_0$ in $w$ is the separator $\#$,
as shown in Figure~\ref{f:theorem_2_pow_n_u3}.
Once the automaton $A_{n+1}$ steps from the last $\#$ in $w$ to the right,
it arrives to the first symbol of $v_0$ in the state $n$
(by the unique transition to the right at $\#$).
The string $v_0$ cannot be empty, because $n \neq 1$.
Once the automaton is inside $v_0$,
it cannot return to $\#$ anymore,
since it has already used the only transition to the right from $\#$,
and cannot use it again without looping.
Therefore, the automaton $A_{n+1}$ starts on the string
$v_0 \in (\Sigma_{n+1}\setminus\{\#\})^*$ in the state $n$,
and, operating as $A_n$,
eventually leaves this string to the right in the state $1$.
Then $|v_0| \geqslant |w_n|$ by the induction hypothesis,
and hence $|v| \geqslant |w_n|$.

\begin{figure}[t]
	\centerline{\includegraphics[scale=1]{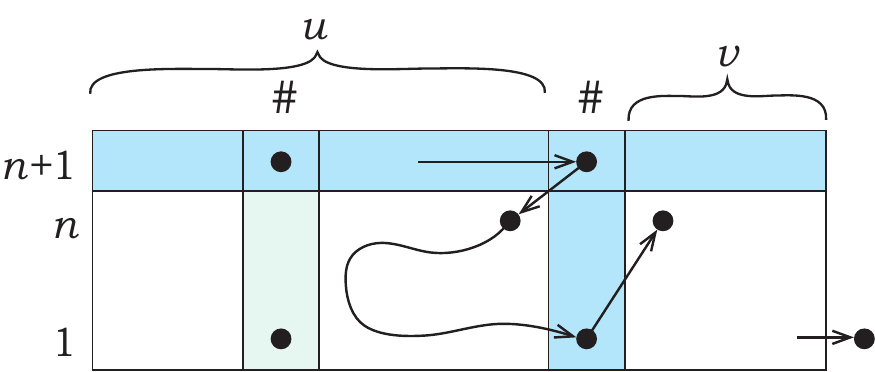}}
	\caption{The case of computations on $u$ not reaching any separators.}
	\label{f:theorem_2_pow_n_u1}
\end{figure}

Now consider the prefix $u$.
Once the automaton $A_{n+1}$ comes in the state $n+1$
to the separator $\#$ between $u$ and $v$,
it moves to the last symbol of $u$ in the state $n$.
In order to leave the string $u$ to the right and proceed further,
it must return to the separator $\#$ in the state $1$,
because there are no transitions by any states $\{2, \ldots, n\}$
at this separator.
If there are no symbols $\#$ in $u$,
or if there are some, but the automaton does not reach them,
then the entire computation of $A_{n+1}$ on $u$
takes place on a certain suffix of $u$ that does not contain $\#$,
as illustrated in Figure~\ref{f:theorem_2_pow_n_u1}.
This computation follows a computation of $A_n$ on a string from $\Sigma_n^*$.
Then, by the induction hypothesis, this suffix is not shorter than $w_n$,
and therefore $|u| \geqslant |w_n|$.

The remaining case is when the automaton
comes to some symbol $\#$ inside the string $u$.
Let $u_0$ be the maximal suffix of $u$ not containing any symbols $\#$,
as in Figure~\ref{f:theorem_2_pow_n_u0}.
The automaton $A_{n+1}$ visits the separator $\#$ to the left of $u_0$,
and then immediately moves from this separator
back to the first symbol of $u_0$ in the state $n$
(the string $u_0$ is non-empty, because it is followed by $\#$, which has no transitions in the state $n$). 
Returning back to $\#$ to the left of $u_0$ is not an option,
since the unique transition by $\#$ to the right has been used already.
Therefore, the automaton leaves $u_0$ by a transition to the right,
and comes to the separator $\#$ between $u$ and $v$.
In order to continue the computation, it should come there in the state $1$.
By the induction hypothesis for this computation on $u_0$,
the length of $u_0$ is at least $|w_n|$.
Then the length of the entire $u$ is also at least $|w_n|$.

\begin{figure}[t]
	\centerline{\includegraphics[scale=1]{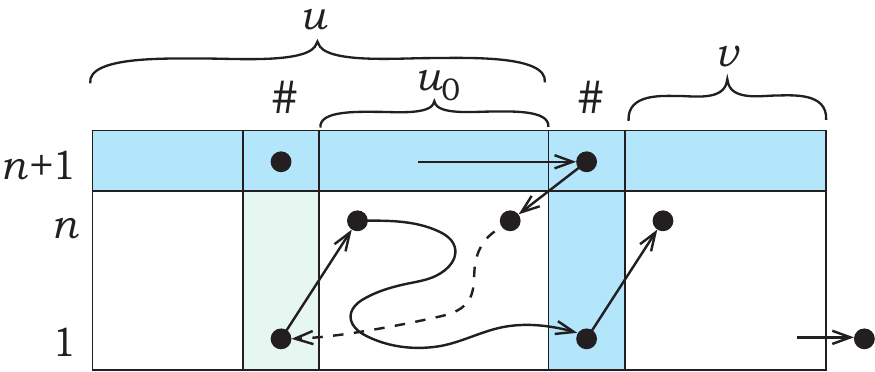}}
	\caption{The case of computations on $u$ reaching a separator $\#$ inside $u$.}
	\label{f:theorem_2_pow_n_u0}
\end{figure}

This confirms that $|w|=|u|+1+|v| \geqslant |w_n|+1+|w_n| = |w_{n+1}|$
and completes the proof.
\end{proof}

\section{Conclusion}

The maximum length of the shortest accepted string
for direction-determinate 2DFA has been determined precisely,
whereas for 2DFA of the general form,
a lower bound of the order $2^n$ has been established.
The known upper bound on this length is of the order $4^n$.
Bounds on the maximum length of shortest strings
for small values of the number of states $n$
are given in Table~\ref{tab:bounds_for_small_N}.

\begin{table}[t]
\begin{center}
\begin{tabular}{|c|r|r|r|r|}
\hline
\multirow{3}{*}{$n$}
& direction-determinate
& \multicolumn{3}{|c|}{2DFA of the general form}
\\
\cline{3-5}
& 2DFA
& lower bound
& computed values
& upper bound \\
& $\binom{n}{\lfloor n/2 \rfloor} - 1$
& $3 \cdot 2^{n-2} - 1$
&
& $\binom{2n}{n+1} - 1$ \\
\hline
2 & 1 & 2 & 2 & 3 \\
\hline
3 & 2 & 5 & 6 & 14 \\
\hline
4 & 5 & 11 & 17 & 55 \\
\hline
5 & 9 & 23 & 32 & 209 \\
\hline
6 & 19 & 47 & & 791 \\
\hline
\end{tabular}
\end{center}
\caption{The maximum length of shortest accepted strings for $n$-state 2DFA, for small $n$.}
\label{tab:bounds_for_small_N}
\end{table}

\begin{figure}[t]
	\centerline{\includegraphics[scale=1]{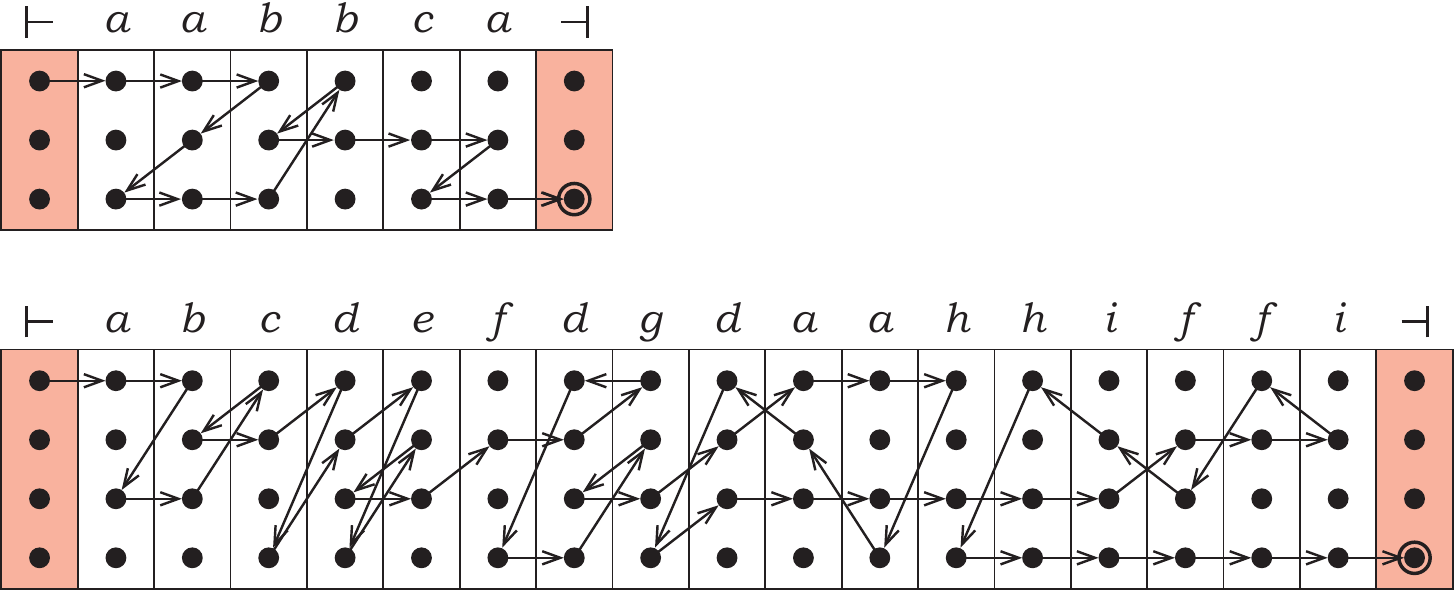}}
	\caption{Automata found by computer programs,
		and their shortest strings:
		(top) 3 states, string of length 6;
		(bottom) 4 states, string of length 17.}
	\label{f:2dfa_shortest_calc_3_4}
\end{figure}

In the table, besides the theoretical bounds,
there are also some computed values
of the length of shortest strings in some automata.
The example for $n=3$ was obtained by exhaustive search,
while the examples for $n=4$ and $n=5$ were found by heuristic search.
Therefore, the maximum length of the shortest string
for 3-state automata is now known precisely,
for 4-state automata it is at least 17 and possibly more,
and the given length of strings for 5 states is most likely much less than possible.
The computations of the automata found for $n=3$ and $n=4$ on their shortest strings
are presented in Figure~\ref{f:2dfa_shortest_calc_3_4}.

It should be noted that these computed values
exceed the theoretical lower bound $\frac{3}{4} \cdot 2^n - 1$ proved in this paper,
and are much less than the known upper bound $\binom{2n}{n+1} - 1$.
Thus, the bounds for 2DFA of the general form are still in need of improvement.

\end{document}